\documentclass[%
 reprint,showpacs,showkeys,nofootinbib,amsmath,amssymb, aps,]{revtex4-1}

\usepackage{graphicx}% Include figure files
\usepackage{dcolumn}% Align table columns on decimal point
\usepackage{bm}% bold math

\usepackage[dvips]{color}

\begin{document}

\title{Cosmological constraints on mirror matter parameters.}

\author{Paolo Ciarcelluti}
 \email{paolo.ciarcelluti@gmail.com}
 \affiliation{Web Institute of Physics, www.wiph.org}

\author{Quentin Wallemacq}
 \email{quentin.wallemacq@ulg.ac.be}
 \affiliation{IFPA, D\'ep. AGO, Universit\'e de Li\`ege, 4000 Li\`ege, Belgium}

\date{\today}% It is always \today, today,

\begin{abstract}
Up-to-date estimates of the cosmological parameters are presented as a result of numerical simulations of cosmic microwave background and large scale structure, considering a flat Universe in which the dark matter is made entirely or partly of mirror matter, and the primordial perturbations are scalar adiabatic and in linear regime.
A statistical analysis using the Markov Chain Monte Carlo method allows to obtain constraints of the cosmological parameters.
As a result, we show that a Universe with pure mirror dark matter is statistically equivalent to the case of an admixture with cold dark matter.
The upper limits for the ratio of the temperatures of ordinary and mirror sectors are around 0.3 for both the cosmological models, that show the presence of a dominant fraction of mirror matter, $0.06 \lesssim \Omega_{\rm mirror} h^2 \lesssim 0.12$.
\end{abstract}



\maketitle

%%%%%%%%%%%%%%%%%%%%%%%%%%%%%%%%%%%%%%%%%%%%%%%%%%%%%%%%%%%%%%%%%%%%%%%%%%%%%%%%
\section{Introduction}
Since a missing mass in the Universe was pointed out in the 1930s, astrophysical evidences for dark matter have been accumulating, increasingly confirming its presence at all cosmological scales. 
But even if the processes of structure formation can draw a picture of its main features, the elementary nature of dark matter remains unknown.

Modern cosmology provides powerful tools for testing dark matter: Big Bang Nucleosynthesis (BBN), the Cosmic Microwave Background (CMB) and the Large Scale Structure (LSS) power spectra can be reproduced in numerical simulations to better discriminate between the different classes of candidates, one of them giving good agreements with all these observations being known as Cold Dark Matter (CDM).
Similarly, mirror dark matter can also account for the cosmological observations, and it is of primary interest to determine whether or not it gives a better agreement than CDM, or if it is equivalent.

Mirror matter was originally proposed by Lee and Yang~\cite{Lee:1956qn}, and further considered by several authors~\cite{Blinnikov:1983gh,Kolb:1985bf,Khlopov:1989fj,Foot:1991bp}, in order to restore the parity symmetry of the Lagrangian of the Standard Model. 
The most natural way to do so is to add to the existing Lagrangian its parity-symmetric counterpart, so that the whole Lagrangian is invariant under the parity transformation, each part transforming into the other. 
This corresponds to reintroduce all the known fields with the same coupling constants, but with opposite parities. 
We therefore end up with a new sector of particles, called mirror sector, which is an exact duplicate of the ordinary sector, but where ordinary particles have left-handed interactions, mirror particles have right-handed interactions~\cite{Foot:1991bp}.

As a consequence, the three gauge interactions act separately in each sector, the only link between them being gravity. 
Because mirror baryons, just like their ordinary counterparts, are stable and can be felt only through their gravitational effects, the mirror matter scenario provides an ideal interpretation of dark matter. 
Its particularity is that it is a self-interacting candidate\footnote{
Astrophysical constraints on self interactions of dark matter present in literature are valid only for homogeneous distributions of dark matter particles, and are therefore not directly applicable to the mirror matter case.}, but without any new parameter at the level of particle physics.

Since it was introduced, mirror matter has been widely studied and its compatibility with experimental and observational constraints has been verified~\cite{Ciarcelluti:2010zz,Okun:2006eb,Ciarcelluti:2003wm,Foot:2003eq,Ciarcelluti:2008qk,Ciarcelluti:2010dm}. Some potential indications of its existence came from the observation of neutron stars~\cite{Sandin:2008db,Ciarcelluti:2010ji}, but also from the direct-search experiments, for which mirror matter gives one of the few possible explanations~\cite{Foot:2008nw,Foot:2012rk}. 
Previous analytical and numerical studies on CMB and LSS power spectra~\cite{Ciarcelluti:2004ip,Ciarcelluti:2004ik,Ciarcelluti:2003wm,Berezhiani:2003wj,Ciarcelluti:2004ij,Ignatiev:2003js} have given encouraging results, but have only limited the parameter space of mirror dark matter. 
Here, we explore it in detail using the fast numerical code CAMB, in order to quantify the compatibility of mirror matter with cosmological observations and to obtain constraints on its parameters.

In Section \ref{section2}, we recall the important cosmological quantities and epochs in presence of mirror matter. 
We describe briefly how we modified the numerical codes CAMB and cosmoMC in order to incorporate the mirror components into the evolution of the Universe in Section \ref{section3}. 
We then present, in Section \ref{section4}, the 1-$\sigma$ constraints on the cosmological parameters coming from the CMB and LSS data for different dark matter compositions, i.e. pure CDM, mixed CDM and mirror matter, and pure mirror matter, and we compare the corresponding best-fit models.
We show also the temporal evolution of the density perturbations for the different components of the Universe.

%-------------------------------------------------------------------------------

\section{Basic cosmology with mirror matter} \label{section2}
Even if the ordinary and mirror sectors are characterized by identical Lagrangians, and obey therefore the same physical laws, their macroscopic realizations are not necessarily the same. 
The differences between the evolutions of the two sectors are parametrized by two cosmological free parameters: the ratio $ x $ of the temperatures of the ordinary and mirror cosmic background radiations; the relative amount $ \beta $ of mirror baryons compared to the ordinary ones.
\begin{eqnarray}\label{mir-param}
x \equiv \left( \frac{S'}{S} \right)^{1/3} \simeq \frac{T'}{T}
~~~~~~ {\rm and} ~~~~~~
\beta \equiv \frac{\Omega'_{\rm b}}{\Omega_{\rm b}} ~~,
\end{eqnarray}
where $T$ ($T'$), $\Omega_{\rm b}$ ($\Omega'_{\rm b}$), and $S$ ($S'$) are respectively the ordinary (mirror) photon temperature, cosmological baryon density (normalized to the critical density of the Universe), and entropy per comoving volume~\cite{Ciarcelluti:2010zz}.

There are several components that contribute to the total present energy density $\Omega_{\rm tot}$: the energy density of relativistic species (radiation) $ \Omega_{\rm r} $, the energy density of non-relativistic species (matter) $ \Omega_{\rm m} $, and the energy density of the vacuum (cosmological constant or dark energy) $ \Omega_\Lambda $. According to the observations of the CMB anisotropies, $\Omega_{\rm tot}=\Omega_{\rm m} + \Omega_{\rm r} + \Omega_\Lambda \approx 1$, meaning that the Universe today is almost flat. In presence of mirror matter, mirror components are present in both radiation and matter energy densities, and the latter is expressed by
\begin{equation}
\Omega_{\rm m} = \Omega_{\rm b}+\Omega'_{\rm b}+\Omega_{\rm DM}
         = \Omega_{\rm b} (1+\beta) + \Omega_{\rm DM}\;,
\end{equation}
where the term $\Omega_{\rm DM}$ includes the contributions of any other possible dark matter particles but mirror baryons.

It can be shown that during BBN, the mirror species $\gamma'$, ${e^{\pm}}'$ and $\nu'_{e,\mu,\tau}$, respectively for mirror photon, electrons, positrons and neutrinos, bring a contribution to the relativistic degrees of freedom equivalent to an effective number of extra neutrino families $\Delta N_\nu 
%= N_\nu -3 = (10.75/1.75) 
\simeq 6.14\,x^4$.
In view of the current bounds on $\Delta N_\nu$~\cite{Izotov:2010ca}, this corresponds to an upper limit $x \lesssim 0.7$, which means that the temperature of the mirror sector was smaller than that of the ordinary one at the epoch of nucleosynthesis, $T'<T$.

Due to the separate conservation of entropies in the two sectors, this initial temperature difference holds throughout the expansion of the Universe, so that the cosmological key epochs take place at different redshifts in the two sectors, happening earlier in the mirror sector than in the ordinary one ~\cite{Ciarcelluti:2004ik,Ciarcelluti:2010zz}. 
The relevant epochs for structure formation are the matter-radiation equality (MRE), the matter-radiation decouplings (MRD) in the ordinary and mirror sectors, and the photon-baryon equipartitions in each sector, occurring respectively at the redshifts  $z_{\rm eq}$, $z_{\rm dec}$, $z'_{\rm dec}$, $z_{\rm b\gamma}$ and $z'_{\rm b\gamma}$.

The MRE is common to both sectors and, in presence of mirror matter, happens at the redshift
\begin{equation} \label{z-eq} 
1+z_{\rm eq}= {\frac{\Omega_{\rm m}}{\Omega_{\rm r}}} \approx 
 2.4\cdot 10^4 {\frac{\Omega_{\rm m}h^2}{1+x^4}} \;,
\end{equation}
while the MRDs and photon-baryon equipartitions in each sector are respectively related by
\begin{equation} \label{z'_dec}
1+z'_{\rm dec} \simeq x^{-1} (1+z_{\rm dec}) 
\;, ~~
\end{equation}
and
\begin{equation} \label{shiftzbg} 
1+z_{\rm b\gamma}' 
  = {\frac{\Omega_{\rm b}'}{\Omega_{\gamma}'} } 
  \simeq \frac{ \Omega_{\rm b} \, \beta}{\Omega_\gamma \, x^4} 
  = (1+z_{\rm b\gamma}) { \frac{\beta}{x^4} } > 1+z_{\rm b\gamma} \;,
\end{equation}
since $T'_{\rm dec} \simeq T_{\rm dec}$ up to small corrections to Eq.~(\ref{z'_dec}). 
Because $x>0$, the value of $z_{\rm eq}$ obtained in the mirror scenario is always smaller than in the standard case, while the upper bound  $x \lesssim 0.7$ ensures that $z'_{\rm dec}>z_{\rm dec}$ and $z'_{\rm b\gamma}>z_{\rm b\gamma}$, showing that the MRD and the photon-baryon equipartition occur earlier in the mirror sector.

By identifying $z_{\rm eq}$ and $z'_{\rm dec}$ from Eqs.~(\ref{z-eq}) and (\ref{z'_dec}), one obtains a reference value $x_{\rm eq}$~\cite{Ciarcelluti:2004ik,Ciarcelluti:2010zz} under which the mirror photons decouple during the radiation-dominated era, with the consequence that the primordial perturbations evolve, in the linear regime, in a way that is very similar to the standard CDM case. Also, analytical and numerical studies on CMB and LSS power spectra~\cite{Ciarcelluti:2004ip,Ciarcelluti:2004ik,Ciarcelluti:2003wm,Berezhiani:2003wj,Ciarcelluti:2004ij,Ignatiev:2003js,Foot:2012ai} have already shown, by comparing qualitatively the results with the observations, that for relatively cold mirror sectors ($x\lesssim0.3$), the dark matter of the Universe can be fully realized by mirror baryons, while for higher $x$ ($x\gtrsim0.3$) mirror baryons and CDM would form an admixture.

%-------------------------------------------------------------------------------

\section{The modified numerical tools CAMB and cosmoMC} \label{section3}
We modified the publicly available code CAMB~\cite{Lewis:1999bs} for the simulation of the anisotropies of the microwave background and the large scale structure of the Universe, together with its Markov Chain Monte Carlo sampling tool cosmoMC~\cite{Lewis:2002ah}, in order to include the mirror components. 
CAMB is used in its most standard mode, i.e. with adiabatic initial conditions for the perturbations, in linear regime, in a flat Universe ($\Omega_{\rm tot}=1$), with an equation of state of vacuum $p_{\Lambda}=-\rho_{\Lambda}$ ($w_{\Lambda}=-1$), without any massive neutrinos and with the standard number of neutrino families $N_{\rm eff}=3.046$.

We defined the necessary mirror variables and included them into the calculations of all the relevant quantities related to the evolution of the background. 
In particular, we considered both ordinary and mirror matter when describing the gravitational interactions. 
As the mirror particles obey the same physical laws as ordinary ones, we doubled the equations related to the evolution of the perturbations of the background, adding the corresponding variables for the perturbations of mirror energy densities and peculiar velocities.

The recombinations are calculated separately in each sector, using two times per model the same code RECFAST~\cite{Seager:1999bc} present in CAMB. This considerably increases the computational time of a model, especially for small values of $x$. 
Indeed, the recombination of mirror hydrogen (as well as of mirror helium) scales as $x^{-1}$, as stated by Eq.~(\ref{z'_dec}), so that the integration of the equations giving the ionized fractions of mirror hydrogen and helium has to be performed over a wider range of redshifts.
However, the program is still fast enough to calculate the huge number of models required for a Monte Carlo analysis of the parameter space.

Compared with previous numerical studies~\cite{Ciarcelluti:2004ip,Ciarcelluti:2004ik,Ciarcelluti:2003wm,Berezhiani:2003wj,Ciarcelluti:2004ij}, we have used here an updated estimate of the primordial composition of mirror particles from Refs.~\cite{Ciarcelluti:2008vm,Ciarcelluti:2008vs,Ciarcelluti:2009da,Ciarcelluti:2010zz,Ciarcelluti:2014vta}, checking that the models obtained using CAMB with this more accurate treatment of mirror BBN are consistent with the previous ones.

We performed Markov Chain Monte Carlo (MCMC) analyses of the parameter space constituted by the standard cosmological parameters plus the two mirror ones using cosmoMC. 
To this aim, we added the mirror parameters to the parameter list of cosmoMC and linked them to the equivalent ones in CAMB.
We end up with an eight-dimensional set of cosmological parameters for which we adopt flat priors and broad distributions, as summarized in Table~\ref{range-par}. 
$\Omega_{\rm b} h^2$ and $\Omega_{\rm cdm} h^2$ are respectively the baryon and cold dark matter densities, $x$ and $\beta$ are the mirror photon temperature and mirror baryon density relative to the corresponding ordinary quantities, $\theta_{\rm s}$ is the ratio of the sound horizon to the angular diameter distance at decoupling, $\tau$ is the reionization optical depth, $n_s$ is the scalar spectral index and $A_s$ is the scalar fluctuation amplitude.

%---------------------------------------------------------

\begin{table}[h]
\caption{\label{range-par}
Adopted flat priors for the parameters.}
\begin{ruledtabular}
\begin{tabular}{ccc}
\textrm{parameter} & \textrm{lower limit} & \textrm{upper limit} \\
\colrule
$\Omega_{\rm b} h^2$     & 0.01 & 0.1\\
$\Omega_{\rm cdm} h^2$ & 0.01 & 0.8\\
$x$ & 0.05 & 0.7\\
$\beta$ & 0.5 & 9.0\\
100 $\theta_{\rm s}$ & 0.1 & 10\\
$\tau$ & 0.01 & 0.8\\
$n_{\rm s}$ & 0.7 & 1.3\\
$\ln(10^{10}A_{\rm s})$ & 2.7 & 4
\end{tabular}
\end{ruledtabular}
\end{table}

The upper limit on $x$ is set by the BBN limit mentioned in Section \ref{section2}. 
CAMB also calculates derived parameters such as the matter and dark energy densities $\Omega_{\rm m}$ and $\Omega_{\Lambda}$, the redshift of reionization $z_{\rm re}$, the Hubble parameter $h$, the age of the Universe in Gyr and the density fluctuation amplitude $\sigma_8$ at $8 h^{-1} {\rm Mpc}$. 
The runs include, as in the standard version of CAMB, weak priors on the Hubble parameter, $0.4 \le h\le 1.0$, and on the age of the Universe, $10 \le {\rm age (Gyr)} \le 20$.

%-------------------------------------------------------------------------------

\section{Results} \label{section4}

\begin{figure}[t]
\includegraphics[scale=0.52]{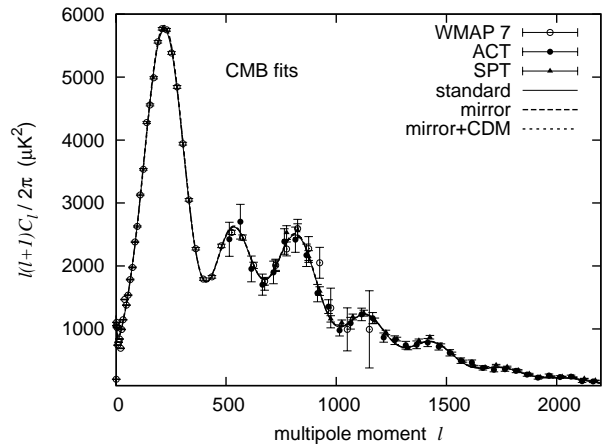}
\caption{\label{fits_CMB}  CMB power spectrum for best-fit models with baryons and mirror matter (dashed line), or baryons, mirror matter, and cold dark matter (dotted line) obtained using CMB only data.
For comparison we show also the standard model fit (solid line).}
\end{figure}

\begin{figure}[t]
\includegraphics[scale=0.52]{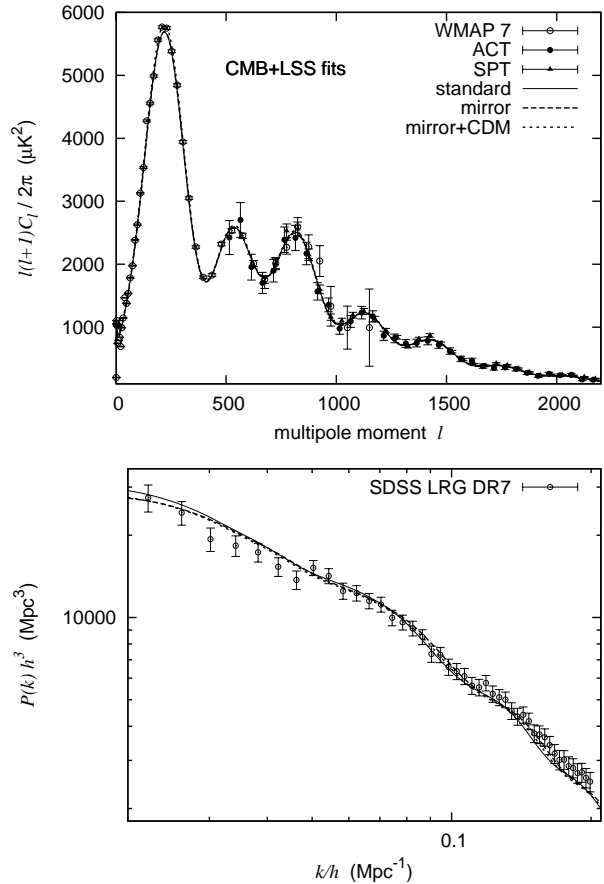}
\caption{\label{fits_CMBLSS} CMB and LSS power spectra for best-fit models with baryons and mirror matter (dashed line), or baryons, mirror matter, and cold dark matter (dotted line) obtained using both CMB and LSS datasets.
For comparison we show also the standard model fit (solid line).}
\end{figure}

This modified version of CAMB can be used in three configurations of dark matter: standard CDM, pure mirror matter, and mixed mirror-CDM matter. 
We considered all these three cases and performed for each of them two analyses, one using the CMB data only and the other using the CMB data combined with the LSS ones. 
The standard CDM case, with the same assumptions and priors, serves as a reference model.

The CMB datasets come from the WMAP7 team~\cite{Larson:2010gs}, together with the APT~\cite{Sievers:2013wk} and SPT~\cite{Keisler:2011aw} observations. 
The former provides the acoustic oscillations of the Cosmic Microwave Background on degree scales with limited cosmic-variance precision while the other two give accurate power spectra at higher {\sl l}'s. 
For the LSS, we included the power spectrum from the SDSS-DR7 luminous red galaxy sample~\cite{Reid:2009xm} limited to $k \lesssim 0.2 h\:{\rm Mpc^{-1}}$, corresponding to sufficiently large length scales, where non-linear clustering and scale-dependent galaxy biasing don't have to be taken into account.

In Table~\ref{params}, we show the 1-$\sigma$ constraints on the parameters obtained from the different dark matter compositions and cosmological tests. 
The density probability of the parameter $x$ was found to be almost flat in the low-$x$ region and sharply decreasing at higher $x$. 
For that reason, we chose to give the upper limits on that parameter at the $95\%$ confidence level. 
The best-fit models obtained by using the CMB data only and both CMB and LSS data are shown in Figs.~\ref{fits_CMB} and \ref{fits_CMBLSS} respectively.

\begin{table*}[t]
\caption{\label{params}1-$\sigma$ constraints on the parameters obtained using different dark matter compositions and cosmological tests.
For the parameter $x$ we reported the upper limit computed at the 95\% c.l.}
\begin{ruledtabular}
\begin{tabular}{ccccccc}
{\sl parameter}        & standard & standard & mirror & mirror  & mirror+CDM & mirror+CDM \\
                       & CMB      & CMB+LSS  & CMB    & CMB+LSS & CMB        & CMB+LSS    \\
\hline
{\sl primary}&&&&&& \\
\hline
$\Omega_{\rm b} h^2   $&$0.02213\pm0.00041$&$0.02205\pm0.00034$&$0.02213\pm0.00040$&$0.02215\pm0.00045$&$0.02225\pm0.00044$&$0.02201\pm0.00034$\\
$\Omega_{\rm cdm} h^2 $&$0.1113 \pm0.0046 $&$0.1161 \pm0.0031 $&$ -               $&$ -               $&$0.026  \pm0.012  $&$0.036  \pm0.010  $\\
$n_{\rm s}            $&$0.9616 \pm0.0097 $&$0.9578 \pm0.0081 $&$0.966  \pm0.011  $&$0.961  \pm0.013  $&$0.964  \pm0.012  $&$0.9558 \pm0.0061 $\\
$\ln(10^{10}A_{\rm s})$&$3.051  \pm0.024  $&$3.074  \pm0.021  $&$3.082  \pm0.034  $&$3.090  \pm0.032  $&$3.100  \pm0.038  $&$3.072  \pm0.018  $\\
$100\,\theta_{\rm s}  $&$1.0406 \pm0.0017 $&$1.0404 \pm0.0015 $&$1.0413 \pm0.0016 $&$1.0403 \pm0.0016 $&$1.0408 \pm0.0017 $&$1.0400 \pm0.0014 $\\
$\tau                 $&$0.073  \pm0.012  $&$0.075  \pm0.011  $&$0.083  \pm0.014  $&$0.085  \pm0.016  $&$0.089  \pm0.016  $&$0.0755 \pm0.0085 $\\
\hline
{\sl mirror}&&&&&& \\
\hline
$x        $&$ -               $&$ -               $&$<0.456           $&$<0.297    
          $&$<0.479           $&$<0.315           $\\
%$x\; (95\%)           $&$ -               $&$ -               $&$<0.496 (0.255)   $&$<0.297 (0.227)   $&$<0.479 (0.269)   $&$<0.315 (0.242)   $\\
$\beta                $&$ -               $&$ -               $&$5.13   \pm0.30   $&$5.21   \pm0.21   $&$4.03   \pm0.50   $&$3.64   \pm0.46   $\\
\hline
{\sl derived}&&&&&& \\
\hline
$\Omega_{\rm m}       $&$0.267 \pm0.025   $&$0.292  \pm0.017  $&$0.273  \pm0.031  $&$0.285  \pm0.020  $&$0.285  \pm0.030  $&$0.291  \pm0.017  $\\
$\Omega_\Lambda       $&$0.733 \pm0.025   $&$0.708  \pm0.017  $&$0.727  \pm0.031  $&$0.715  \pm0.020  $&$0.715  \pm0.030  $&$0.709  \pm0.017  $\\
$z_{\rm re}           $&$9.3   \pm1.0     $&$9.70   \pm0.95   $&$10.3   \pm1.2    $&$10.5   \pm1.3    $&$10.8   \pm1.4    $&$9.73   \pm0.77   $\\
$h                    $&$0.710 \pm0.022   $&$0.690  \pm0.013  $&$0.708  \pm0.024  $&$0.695  \pm0.017  $&$0.698  \pm0.023  $&$0.690  \pm0.014  $\\
age [Gyr]              &$13.750\pm0.092   $&$13.793 \pm0.066  $&$13.687 \pm0.093  $&$13.759 \pm0.088  $&$13.71  \pm0.10   $&$13.782 \pm0.067  $\\
$\sigma_8             $&$ -               $&$0.824  \pm0.015  $&$ -               $&$0.767  \pm0.021  $&$ -               $&$0.746  \pm0.018  $\\
\end{tabular}
\end{ruledtabular}
\end{table*}

From Table~\ref{params}, we directly note that the primary cosmological parameters, except the CDM energy density, do not vary significantly from one kind of model to the other and for both analyses (CMB and CMB+LSS). 
On the other hand, some derived parameters are more perturbed, as the total matter density, that increases at the expenses of the dark energy density in models using CMB data only. 
This is coupled to the decrease of the Hubble parameter, which still falls within the current constraints. 
The non-baryonic matter density is in all cases $5$ or $6$ times larger than the baryonic density, as usually derived from standard analyses. 
Turning to the mirror parameters, the $95\%$ c.l. upper limits on $x$ are found to be $x<0.456$ for a pure mirror Universe and $x<0.479$ for the mixed mirror-CDM scenario, in case of the CMB only analysis. 
Adding the LSS constraints in the computations significantly lowers these upper limits, since we obtain $x<0.297$ and $x<0.315$ at $95\%$ c.l. respectively for the pure mirror and mixed mirror-CDM cases. 
This confirms the higher sensitivity on $x$ of the formation of the large scale structure of the Universe already highlighted in previous studies~\cite{Ciarcelluti:2004ip,Ciarcelluti:2010zz}. 
Note that all these allowed intervals for $x$ contain the values that make possible the interpretation of the direct-dark-matter-search experiments by mirror matter~\cite{Foot:2008nw,Foot:2012rk}. 
Finally, the parameter $\beta$, quantifying the presence of mirror matter in the Universe, has similar values with CMB only and CMB combined with LSS. 
In the pure mirror case, $\beta$ lies between $5$ and $5.5$, indicating that mirror models require consistent amounts of mirror matter in order to reproduce the observables. 
In the mixed mirror-CDM case, we find mirror matter densities that are between $2$ and $4$ times larger than those of CDM. 
This suggests that, in a Universe with a dark matter composed of several components, the mirror one would be the dominant part.

The likelihoods of the best-fit models for the three compositions of dark matter have, for each analysis (CMB and CMB+LSS), very close values. 
Respectively for pure CDM, pure mirror and mixed mirror-CDM, they are $-\ln({\cal L})=3772$, $-\ln({\cal L})=3771$ and $-\ln({\cal L})=3771$ in case of the CMB only analysis, and $3795$, $3794$ and $3795$ in the other case. 
In view of the difference of one or two degrees of freedom between the models, these values do not show any statistical preference.

\begin{figure*}
\begin{centering}
\includegraphics[scale=0.53]{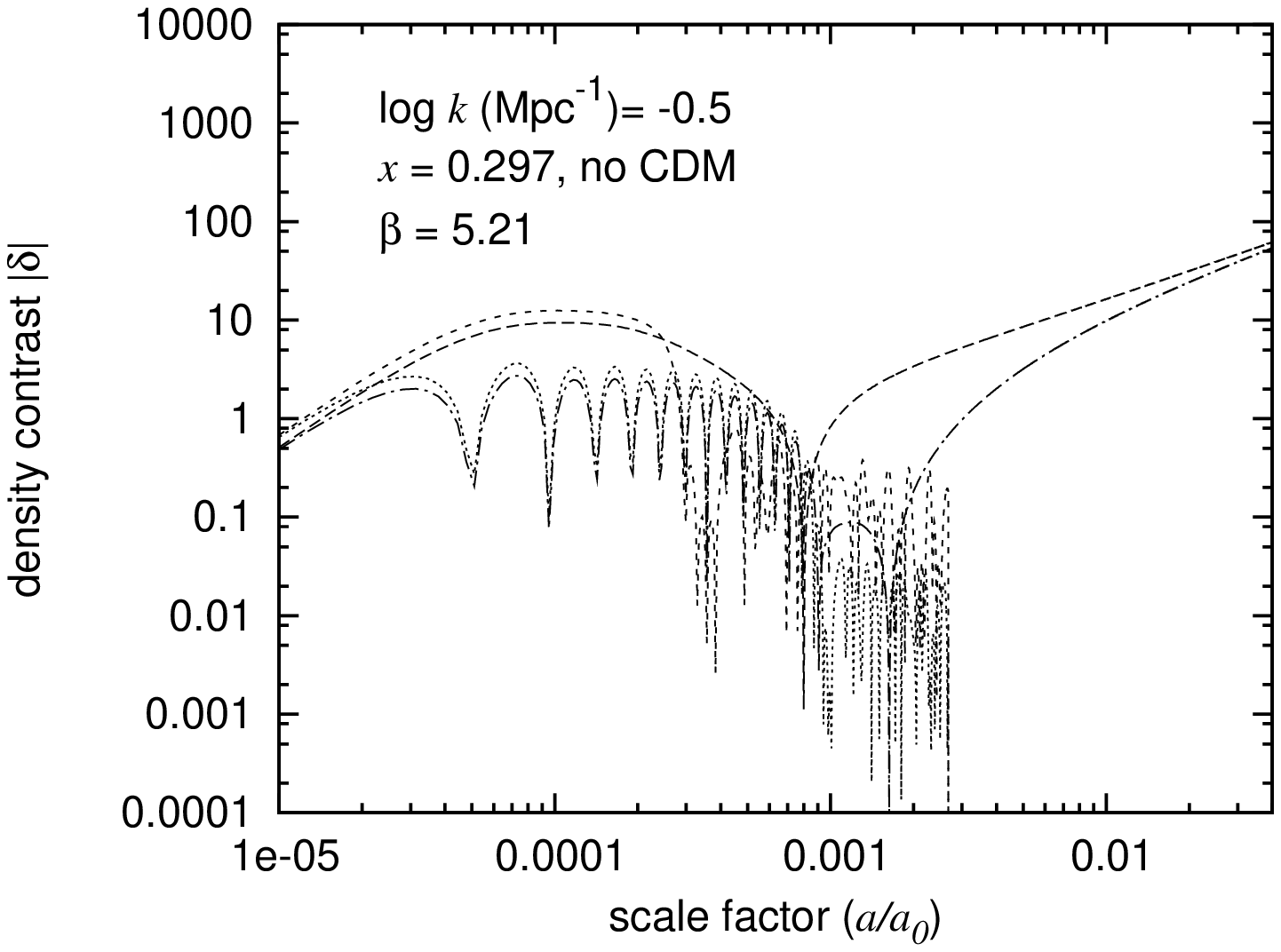}\includegraphics[scale=0.53]{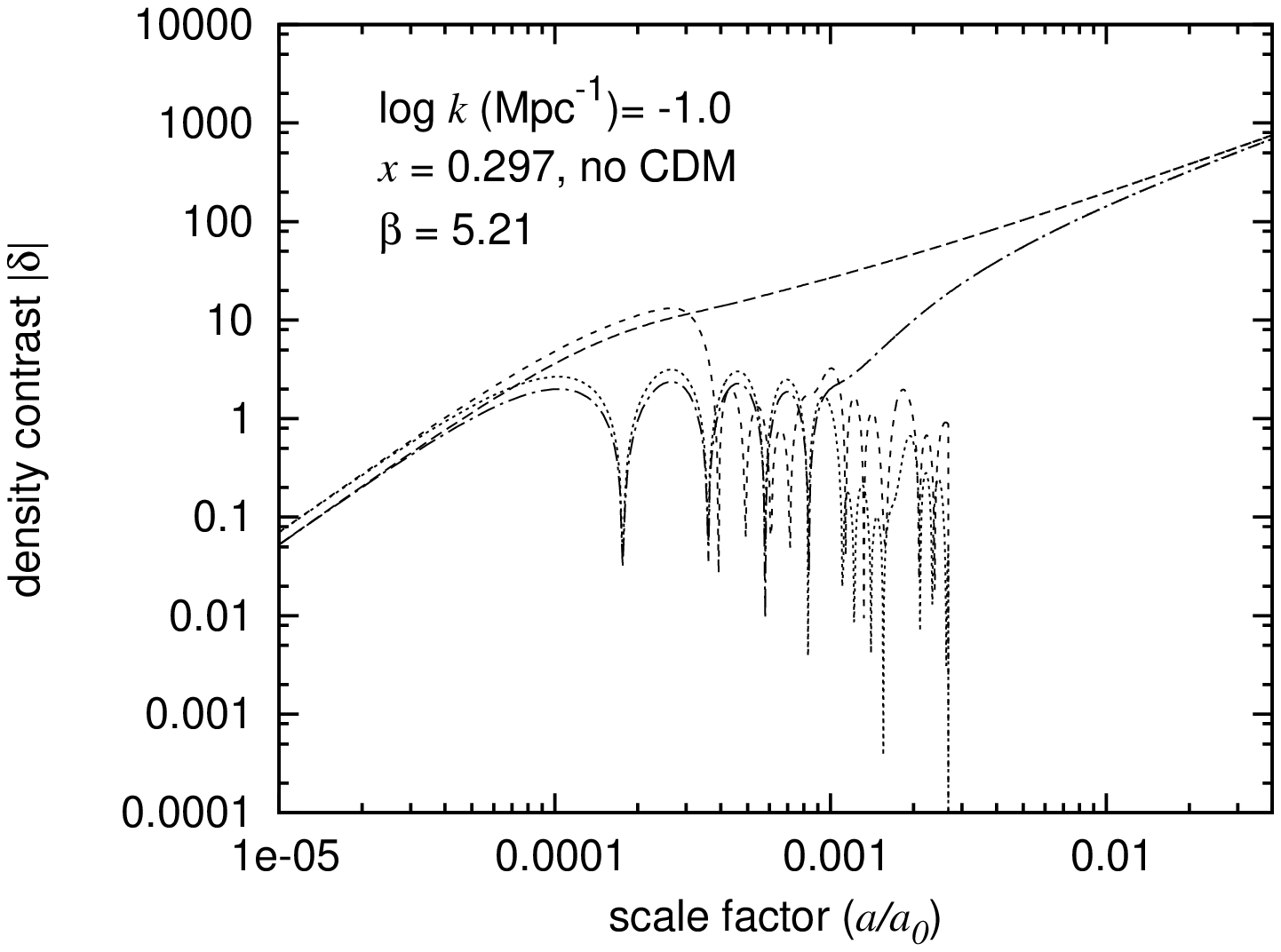}
\par\end{centering}
\begin{centering}
\includegraphics[scale=0.53]{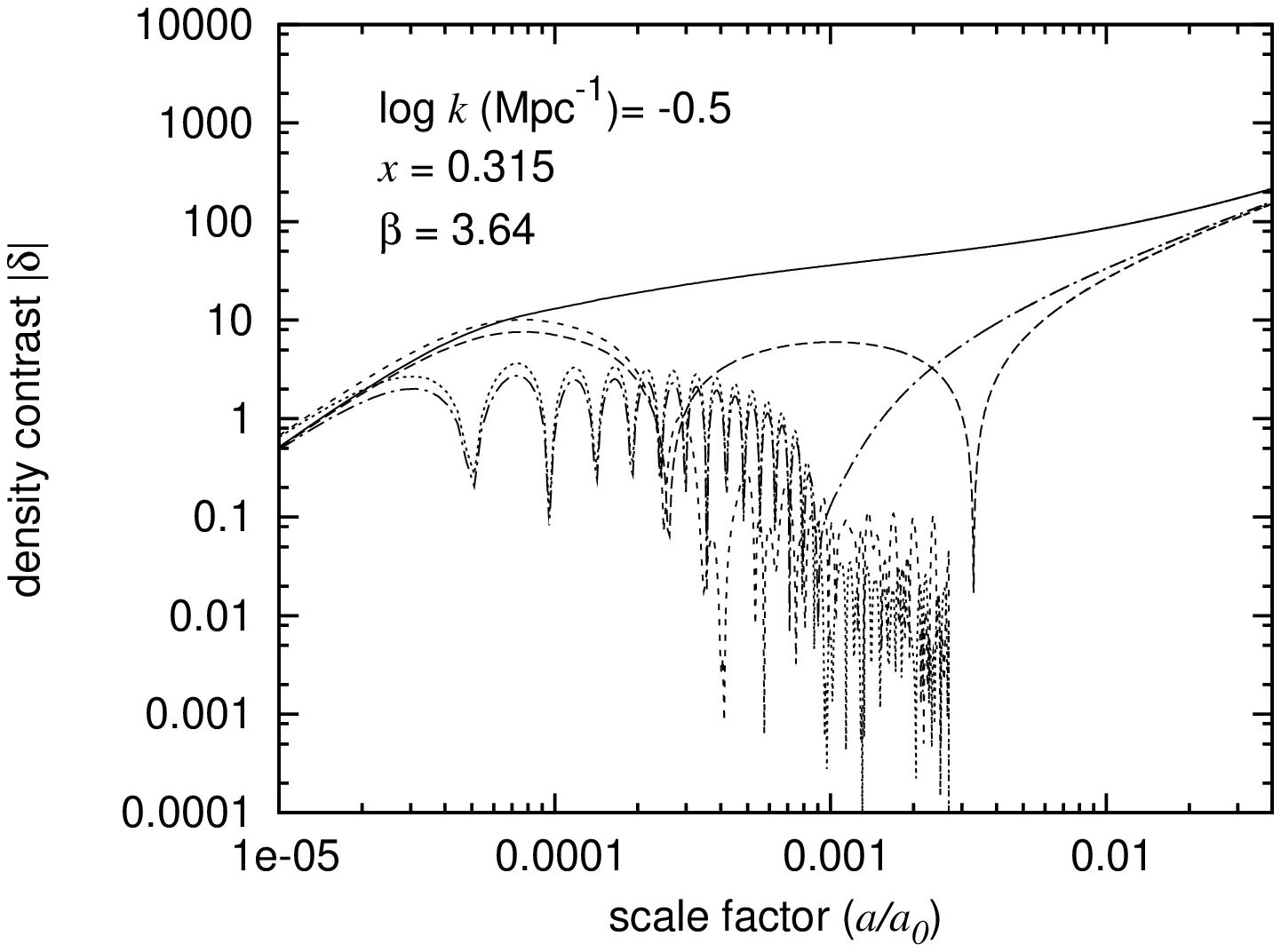}\includegraphics[scale=0.53]{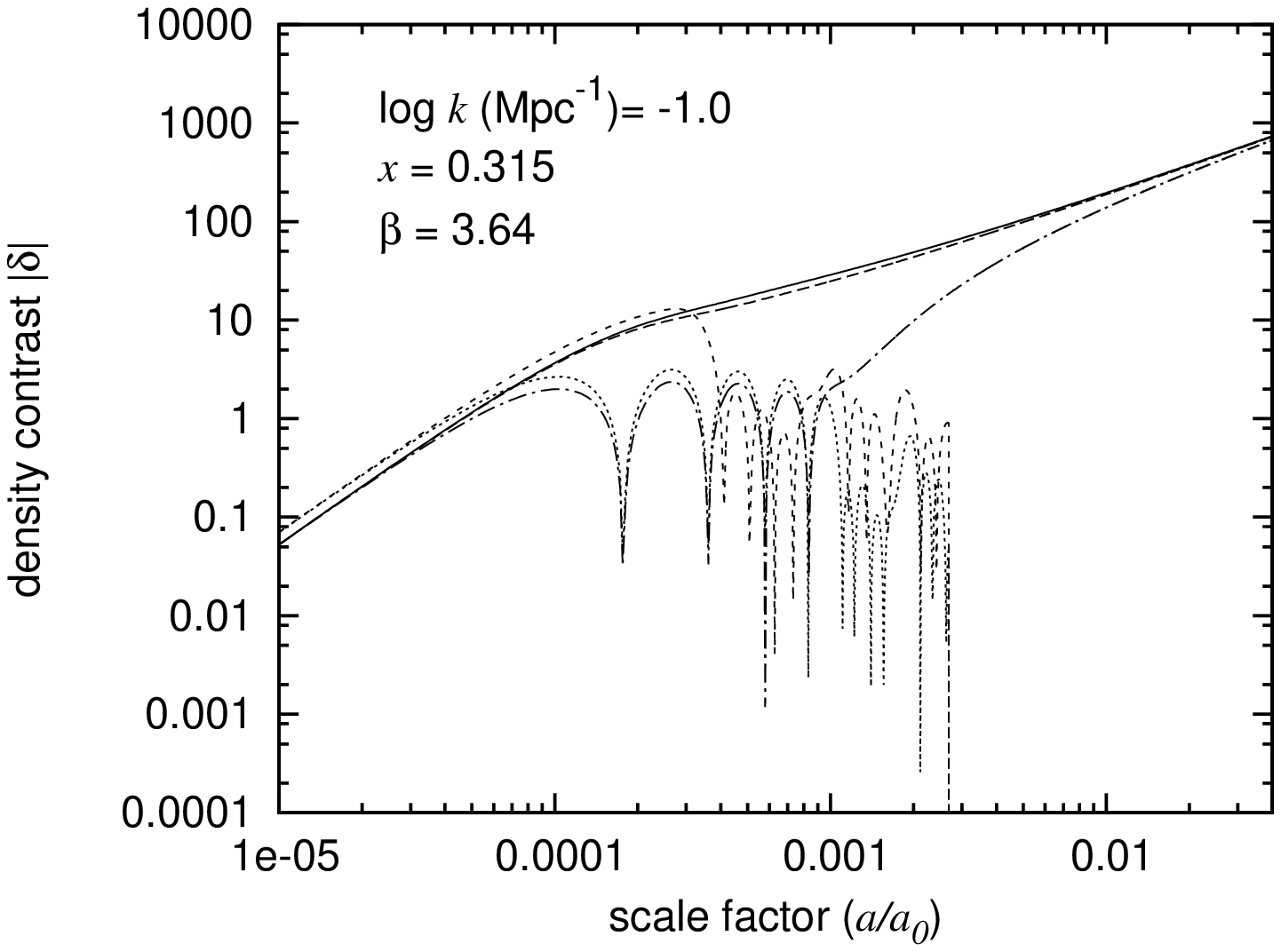}
\par\end{centering}
\caption{\label{evolution}Evolution of the density perturbations of the different components of a mirror Universe: cold dark matter (solid), ordinary baryons (dot-dashed), ordinary photons (dotted), mirror baryons (long dashed) and mirror photons (dashed). 
The models correspond to the parameters of Table~\ref{params} obtained with the CMB and LSS data. 
Top left: pure mirror case at the scale $\log k$(Mpc$^{-1}$)$=-0.5$. 
Top right: pure mirror case at the scale $\log k$(Mpc$^{-1}$)$=-1.0$. 
Bottom left: mixed mirror-CDM case at the scale $\log k$(Mpc$^{-1}$)$=-0.5$. 
Bottom right: mixed mirror-CDM case at the scale $\log k$(Mpc$^{-1}$)$=-1.0$.}
\end{figure*}

In Fig.~\ref{evolution} we show the temporal evolution of density perturbations for CDM, ordinary and mirror baryons, ordinary and mirror photons, computed at two different scales and for our estimated values of the parameters (for $x$ we considered the upper limits) in the two configurations of pure mirror (upper figures) and mixed mirror-CDM (lower figures).
In every plot it is visible that the decoupling of mirror baryons and photons happens before the one of the ordinary species, and this is more evident in the right plots.
In particular, comparing the top and bottom right plots, one can see the equivalence between CDM and mirror matter for this scale of perturbations and this region of $x$.
The left panels are for the perturbations at a smaller scale.
The mentioned CDM-mirror equivalence is at first sight less evident, but considering the final temporal effect (that gives the distribution of the cosmological matter structures that we see today), again the mirror baryons have a role comparable to the CDM, driving the evolutions of the structures, and providing the gravitational seeds where ordinary baryons can fall and accrete.

The fact that we can access only an upper limit on $x$, together with the equivalence of the different best fits, could suggest a mirror matter with a CDM-like behavior. 
Indeed, as announced by Eq.~(\ref{z'_dec}) and further confirmed by Refs.~\cite{Ciarcelluti:2004ik,Ciarcelluti:2004ip,Ciarcelluti:2010zz}, the smaller $x$, the earlier the decoupling of mirror baryons, and the more they behave like cold dark matter at linear scales, the case $x\rightarrow 0$ being equivalent to CDM. 
If such a trend is verified, CDM could find a possible interpretation through mirror matter. 
Future data on LSS, especially in non-linear regimes where mirror matter shows more marked differences with CDM at non-zero $x$, should help to discriminate between mirror and CDM models, or confirm their equivalence.

%-------------------------------------------------------------------------------

\section{Conclusion}
In summary, we presented up-to-date estimates of the cosmological mirror parameters $x$ and $\beta$ coming from the observations of the anisotropies of the Cosmic Microwave Background and of the Large Scale Structure of the Universe. 
The most stringent constraints were obtained by using both the CMB and LSS data, for which we determined that $x<0.297$ ($95\%$ c.l.) and $\beta=5.21\pm0.21$ (1$\sigma$) in the pure mirror case, and $x<0.315$ ($95\%$ c.l.) and $\beta=3.64\pm0.46$ (1$\sigma$) in the mixed mirror-CDM case. 
These parameter ranges contain the values favored by the direct searches for dark matter. 
On the other hand, we have seen that cosmological models with dark sectors constituted by pure CDM, pure mirror matter and both CDM and mirror matter are equivalent concerning the CMB and LSS power spectra. 
The upper limits on $x$ together with the equivalence of the different compositions of dark matter may indicate that, if present, mirror matter could behave like CDM. Future data on LSS at non-linear length scales should help to discriminate between CDM and mirror matter or confirm their equivalence, in which case the latter would be an interpretation of the former.

\begin{acknowledgments}
We are grateful to Jean-R\'en\'e Cudell for useful discussions.
PC acknowledges the hospitality of the IFPA group and the financial support of the Belgian Science Policy during part of this work.
QW is supported by the Belgian Fund FRS-FNRS as a Research Fellow.
\end{acknowledgments}

%%%%%%%%%%%%%%%%%%%%%%%%%%%%%%%%%%%%%%%%%%%%%%%%%%%%%%%%%%%%%%%%%%%%%%%%%%%%%%%%

\bibliography{fit-cosmo-mir-2}% Produces the bibliography via BibTeX.

\end{document}